\documentclass[sigconf]{acmart}
\usepackage{multirow}
\usepackage{tabularx}
\usepackage{xcolor}
\usepackage{array}
\usepackage{stfloats}

\AtBeginDocument{%
  \providecommand\BibTeX{{%
    \normalfont B\kern-0.5em{\scshape i\kern-0.25em b}\kern-0.8em\TeX}}}

\setcopyright{acmcopyright}

\copyrightyear{2024}
\acmYear{2024}
\setcopyright{rightsretained}
\acmConference[CHI '24]{Proceedings of the CHI Conference on Human Factors in Computing Systems}{May 11--16, 2024}{Honolulu, HI, USA}
\acmBooktitle{Proceedings of the CHI Conference on Human Factors in Computing Systems (CHI '24), May 11--16, 2024, Honolulu, HI, USA}
\acmDOI{10.1145/3613904.3641970}
\acmISBN{979-8-4007-0330-0/24/05}

\begin{document}

\title[Reconfigurable Content Tokens for Accessible Data Visualizations]{“Customization is Key”: Reconfigurable Content Tokens for Accessible Data Visualizations}
\author{Shuli Jones}
\email{jonsh@mit.edu}
\orcid{0000-0002-4587-5551}
\affiliation{%
  \institution{Massachusetts Institute of Technology}
  \streetaddress{32 Vassar St}
  \city{Cambridge}
  \state{Massachusetts}
  \country{USA}
  \postcode{02139}
}

\author{Isabella Pedraza Pineros}
\email{ipedraza@mit.edu}
\orcid{0009-0002-3269-7618}
\affiliation{%
  \institution{Massachusetts Institute of Technology}
  \streetaddress{32 Vassar St}
  \city{Cambridge}
  \state{Massachusetts}
  \country{USA}
  \postcode{02139}
}

\author{Daniel Hajas}
\email{d.hajas@ucl.ac.uk}
\orcid{0000-0002-2811-1197}

\affiliation{%
  \institution{Global Disability Innovation Hub}
  \streetaddress{1 Pool Street}
  \city{Stratford}
  \country{UK}
  \postcode{E20 2AF}
}

\author{Jonathan Zong}
\email{jzong@mit.edu}
\orcid{0000-0003-4811-4624}
\affiliation{%
  \institution{Massachusetts Institute of Technology}
  \streetaddress{32 Vassar St}
  \city{Cambridge}
  \state{Massachusetts}
  \country{USA}
  \postcode{02139}
}

\author{Arvind Satyanarayan}
\email{arvindsatya@mit.edu}
\orcid{0000-0001-5564-635X}

\affiliation{%
  \institution{Massachusetts Institute of Technology}
  \streetaddress{32 Vassar St}
  \city{Cambridge}
  \state{Massachusetts}
  \country{USA}
  \postcode{02139}
}

\renewcommand{\shortauthors}{
Jones et al.
}

\begin{abstract}
Customization is crucial for making visualizations accessible to blind and low-vision (BLV) people with widely-varying needs.
But what makes for usable or useful customization?
We identify four design goals for how BLV people should be able to customize screen-reader-accessible visualizations: \textit{presence}, or what content is included; \textit{verbosity}, or how concisely content is presented; \textit{ordering}, or how content is sequenced; and, \textit{duration}, or how long customizations are active.
To meet these goals, we model a customization as a sequence of content tokens, each with a set of adjustable properties.
We instantiate our model by extending Olli, an open-source accessible visualization toolkit, with a settings menu and command box for persistent and ephemeral customization respectively.
Through a study with 13 BLV participants, we find that customization increases the ease of identifying and remembering information. However, customization also introduces additional complexity, making it more helpful for users familiar with similar tools.

\end{abstract}



\maketitle

\section{Introduction}

Customization (or personalization) is a crucial part of digital accessibility, as people who are blind and low-vision (BLV) are a heterogeneous population with widely-varying needs~\cite{morris_rich_2018, stangl_person_2020, elavsky_option-driven_2023}.  
Moreover, as they may have different levels of experience with assistive technologies\,---\,like tactile graphics or screen readers\,---\,BLV people have different preferences and mental models when interacting with digital media~\cite{zong_rich_2022, sharif_understanding_2021}. 
Current approaches to accessible visualization, however, only offer fixed methods of accessibility\,---\,for instance, static natural language descriptions that offer the same experience to all users regardless of their prior experience or preferences \cite{morris_rich_2018, stangl_person_2020, jung_communicating_2021, lundgard_accessible_2022}. 
As a result, even if a given description follows best practices, some BLV users may find the description provides too much detail, burying relevant signal near the end of a lengthy sequence of extraneous information, while others may find that, despite this detail, an important piece of information they rely on is missing~\cite{lundgard_accessible_2022, zong_rich_2022}.

However, it is unclear how best to support customizing accessible visualizations. 
Commonly-used screen readers offer some customization options\,---\,for example, how and whether punctuation like parentheses and brackets should be read out. 
But, these general-purpose options lack the flexibility needed for a specific domain like visualization\,---\,for instance, how should field names be read out (if at all). 
Although there is a growing body of research \cite{zong_rich_2022, joyner_visualization_2022, jung_communicating_2021, kim_accessible_2021, lundgard_accessible_2022, sharif_understanding_2021} and design practice \cite{cesal_writing_2020, elavsky_how_2022} that offers guidance on accessible visualization design, these prescriptions provide general advice that is not able to account for individual differences or preferences\,---\,particularly as these preferences may vary based on the task a BLV user is performing, and over the course of their interactions with a visualization.

To address the diverse information-seeking needs of BLV screen reader users, we conducted an iterative co-design process with Hajas, our blind co-author. 
We identify four design goals for how BLV people should be able to customize screen-reader-accessible visualizations: toggling \textit{presence}, or the content included in the screen reader output; changing \textit{verbosity}, or the length and conciseness of the content; reconfiguring \textit{ordering}, or how the content is sequenced; and setting the \textit{duration}, or for how long a customization is in effect. 
These four goals arise directly from the unique affordances of screen readers: content that is read aloud, in comparison to content that is read with the eyes, takes more time, is harder to skim through, and puts a greater cognitive load on short-term memory \cite{zong_rich_2022, sharif_understanding_2021, ahmed_accessible_2012}.
We develop a conceptual model of a customization as a sequence of individual content tokens, where each content token has its own adjustable properties. 
Breaking content down into tokens allows users to adjust presence by toggling individual tokens on and off, adjust ordering by changing the position of individual tokens in the list, and set verbosity and duration as properties of the tokens or customization respectively.

We instantiate our model in the context of the hierarchical textual descriptions of visualizations produced by Olli, an open-source accessible visualization toolkit~\cite{blanco_olli_2022}.
Thus, a customization is defined as a sequence of \textit{string tokens}, and the content of a token is defined as a function of three parameters: 
\textit{affordance}, which corresponds to the type of task a token supports; \textit{direction}, or how a token refers to information relative to its position in the hierarchy; and \textit{brevity}, which we implement as shorter or longer variations of token text.
We extend the Olli user interface with a \textit{settings menu} to enable specification of persistent customizations, and a \textit{command box} for ephemeral customizations.
Persistent customizations' effects last until they are changed by the user, while ephemeral customizations may revert on their own or last for a single description. Correspondingly, the settings menu is more complex and allows users to build, save, and retrieve customizations, whereas the command box is quick to use and offers a fixed menu of commands that can be executed with just a few keystrokes.

We evaluate our model by conducting a study on Olli with 13 blind and low-vision screen reader users.  We find that users confirm our assumptions that customization crucially supports their autonomy and agency when working with data, but that customization for screen readers is often neglected by interface designers. They also confirm that user preferences and desires around customization vary widely, and that many consider customization interfaces more useful with time and experience. Our model of customization does meet our four design goals, supporting users’ self-guided data exploration by allowing them to access information more efficiently and in ways that suit their tasks and preferences. But even so, customization can still be high-effort to learn and use, underscoring the importance of carefully-designed systems and enough time to build familiarity.

\section{Related Work}

\subsection{Diversity of Screen Reader User Preferences}

The general consensus within the visualization design field is that there is no ``one-size-fits-all'' approach when designing tools to make visualizations meaningful and accessible to screen-reader users \cite{sharif_understanding_2021}. This aligns with disability scholars' critiques of design approaches such as universal design which aim to create non-specialized designs usable by all people \cite{smith_universal_2011}, but which often disregard the significance of discerning and adapting to differences in user needs \cite{hamraie_designing_2013, williamson_getting_2012}. Indeed, sometimes different peoples' needs conflict, and accommodating one need might exacerbate barriers for another \cite{elavsky_option-driven_2023}.

For sighted users, the visualization design field has already identified variations of user tasks, backgrounds, and expertise as necessary considerations in designing more equitable visualizations. Brehmer and Munzner \cite{brehmer_multi-level_2013} define a rich task typology to decompose the different how, why, and what in user interactions with data visualization. Peck et al. \cite{peck_data_2019} draw attention to how the complex tapestry of motivations, preferences, and beliefs of a person can affect their experiences with data visualizations. Stofer and Che \cite{stofer_comparing_2014} found that experts and novices often take very different approaches to viewing and meaning-making when reading visualizations. 

These considerations are further compounded for screen reader users. From a technological standpoint, screen reader users use a diverse set of screen readers and complementary devices \cite{noauthor_screen_nodate} and have different levels of exposure to visualization concepts from non-textual modalities such as tactile graphics. Moreover, screen reader users have diverse backgrounds in education \cite{noauthor_blindness_2019}, internet proficiency, and screen preferences \cite{noauthor_webaim_2018, noauthor_webaim_2021}. Like sighted users, screen reader users employ different kinds of methodologies to learn about visual semantics \cite{potluri_examining_2021}, require different visualization approaches to best suit their needs and expectations \cite{joyner_visualization_2022}, and are likely to engage more with visualizations that communicate data about a topic they're interested in \cite{sharif_understanding_2021}. Unique sets of preferences specific to screen reader users have also emerged; Lundgard and Satyanarayan \cite{lundgard_accessible_2022} identified that the type of semantic content that best communicates a chart's trends and statistics varies between screen reader users and sighted users, as well as among different screen reader users.

As a result of the diversity in screen reader user preferences, researchers have suggested customization around a screen reader user's preferences \cite{lundgard_accessible_2022, morris_rich_2018, jung_communicating_2021} as an important approach in ensuring autonomy for screen reader users with exploring and extracting information from visualizations \cite{sharif_understanding_2021}. 
Mindful of the impact that differing needs and preferences have on screen reader users' interactions with visualizations, our work offers an initial step in designing systems that help users express these differences.

\subsection{Customizable User Interfaces}

Computer scientists have long recognized the importance of software that enable users to ``mold and channel its power to [their] own needs'' \cite{kay_personal_1977}.
Recent work in malleable end user software has sought to empower users to shape and appropriate software to suit their personal and idiosyncratic needs \cite{klokmose_webstrates_2015}, regardless of how those functional needs change over time \cite{borowski_varv_2022}.
Within accessibility, researchers have developed theoretical lenses to think about how systems should adapt to user needs.
Ability-based design advances a vision of systems that adapt to users' abilities, often by measuring their behavior and attempting to infer or anticipate the best adaptation \cite{wobbrock_ability-based_2011}. While this approach helpfully reduces the level of effort required from the user to learn how to articulate their needs legibly to the system, systems cannot always correctly anticipate what users need. To contrast and complement ability-based design, researchers have articulated option-driven design \cite{elavsky_option-driven_2023} as an approach based on providing sensible defaults while enabling users to express their own adaptations through series of options. Option-driven design puts more agency in the hands of users, but trades off increased effort required to configure options. Additionally, designers must be mindful of the fact that increasing the number of options alone does not necessarily improve the accessibility of an inherently inaccessible design \cite{elavsky_option-driven_2023}.
Our approach to customization is influenced by option-driven design, as we offer a gradient of customization methods that trade off effort and granularity. Users can stick to default options, choose from a small menu of preset customizations, or manually set individual customization options.

\section{Background: Hierarchical Textual Descriptions of Visualizations with Olli}
\label{sec:background}

This paper builds closely on two pieces of prior work: Zong, Lee, Lundgard et al. \cite{zong_rich_2022}, which introduces design dimensions for rich screen reader experiences of data visualizations, and Blanco et al. \cite{blanco_olli_2022} which instantiates these dimensions as Olli, an open-source library for producing navigable, hierarchically-structured textual descriptions of data visualizations.
In this section, we aim to provide sufficient background to understand the remainder of this paper. 

\begin{figure*}[b!]
  \centering
  \includegraphics[width=\linewidth]{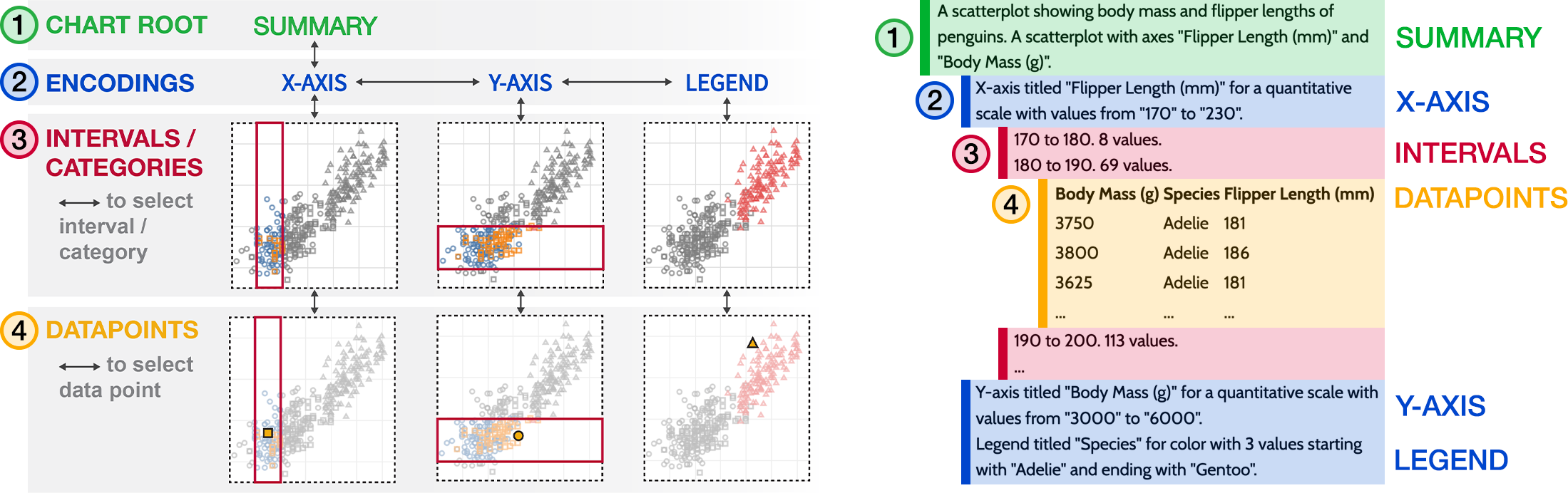}
  \caption{The relationship between the structure of a tree-shaped navigable hierarchy and the visualization's corresponding Olli hierarchy. The left figure is adapted from Zong, Lee, Lundgard et al. \cite{zong_rich_2022}. It shows four levels of the tree and how users can use arrow keys to move between them. The right figure shows how the four levels are instantiated in Olli. Each level of the tree is collapsed by default (in this case the y-axis and legend) and expands as the user moves into the level. This allows users to "zoom in" on selections of the data broken out by any of the three fields (in this case flipper length, body mass, and species).}
  \Description{A graphic with two parts. The left side illustrates an  abstract accessible visualization structure for an example scatterplot, and its corresponding four sections: the Chart Root (i.e., a summary textual description of the chart), Encodings (i.e., the chart’s x-axis, y-axis, and legend), Intervals/Categories (i.e, navigating by selecting axis intervals or data categories), and individual Datapoints. Navigating between these different sections is accomplished with the arrow keys. The right side shows this abstract structure instantiated in Olli for an example scatterplot, and labels which sections of text in Olli correspond to each of the four entities from the left side of the graphic.}
  \label{background}
\end{figure*}

Zong, Lee, Lundgard et al. introduce three design dimensions for enabling rich screen reader experiences of accessible data visualizations: \textit{structure} (how the individual elements of the representation are arranged), \textit{navigation} (how the user moves between elements), and \textit{description} (what is read out at each element). 
Olli is an open-source library that instantiates these design dimensions. It converts web-based visualizations into screen reader accessible representations whose \textit{structure} is that of a tree and where the user can use \textit{structural navigation} to move around the nodes. Node \textit{descriptions} contain content about the chart type, axis labels, and counts of data, implemented as a fixed set of string templates.
The content of the descriptions is produced directly from the underlying dataset for the visualization, providing consistency of text across visualizations regardless of style or chart type (tables, bar and line charts, stacked bar, small multiples, etc.). 

The structure of the hierarchies is produced by the encodings of the visualization. This means that they share a similar format with the original visualization. Each hierarchy has four or five levels. The top-level node (the "summary", or \autoref{background}.1) alerts the user to the existence of the hierarchy and gives an overview of the visualization. If the visualization has multiple facets (for example, a line chart with multiple lines), the next level will break out each facet into its own node. The level after that ("encodings", or \autoref{background}.2) has one node for each axis or legend in the original visualization. Each axis or legend node has multiple children, one for each category in a legend or for each interval of data between the gridlines of an axis ("intervals/categories", or \autoref{background}.3). Finally, the last and most detailed level of the hierarchy is a table containing all of the individual datapoints within the selected interval or category of data ("datapoints", or \autoref{background}.4). This structure, as Zong, Lee, Lundgard et al. found, allows users to get an overview of the visualization first and then the details on demand, with the information at each level of the hierarchy building on the levels above it and filling in additional details for the user. 

The user can move through this hierarchy using their keyboard's arrow keys, or with shortcuts such as \texttt{x} or \texttt{y} that jump directly to the x- and y-axis, respectively. Moving down goes to the next level of detail in the hierarchy, while moving up goes to the previous level. Moving left and right switches between sibling nodes at the same level of detail. For example, here's how a user would access the currently-selected section in \autoref{background}. First, they would start at the root node (1) and learn that they are in a hierarchy for a scatterplot comparing the body mass and flipper length of penguins. Perhaps they want to begin by getting a sense of the distribution of flipper length among penguins. They would press down to move into the axis level, where they would land on the first sibling node, the X-axis, representing flipper length in millimeters (2). Pressing down a second time would move into data sections of the X-axis, in this case sections divided into groups of ten millimeters (3). The first group, 170 to 180 millimeters, has 8 values. If the user presses right they move to the second group, 180 to 190 millimeters, which has 69 values. If the user wanted more information about these values, like which species of penguin they were from, they could press down to enter a table where each datapoint has its own row, each representing one penguin whose flipper length falls between 180 and 190 millimeters (4).
\section{Customizing Screen-Reader-Accessible Visualizations}

Although the approaches described in the previous section begin to provide screen reader users with methods for accessible information-seeking in data visualizations, static textual descriptions cannot support the diverse tasks and preferences of these users. 
To enable screen reader users to reconfigure content as necessary to accomplish their information-seeking task, we introduce a conceptual model for customizing accessible visualizations. We model customizations as compositions of content \emph{tokens} in order to support four design goals we identified: \emph{presence}, or what content is conveyed; \emph{verbosity}, or the length and conciseness of the content's delivery; \emph{ordering}, or the sequencing of tokens used to convey the content; and \emph{duration}, or how long a particular customization lasts.

\subsection{Design Goals}

To identify our design goals, we returned to the studies conducted by Zong, Lee, Lundgard et al. \cite{zong_rich_2022}. In these studies, blind and low-vision (BLV) participants were asked to use tabular and tree-structured representations of visualizations to perform open-ended sensemaking tasks. First, Jones read through and coded the transcripts of these studies, focusing on the \textit{description} dimension of the authors’ design space. For each transcript, Jones noted each time a user felt that a description was or was not meeting their needs. Next, Jones created informal thematic groupings of these instances (for example, \textit{``missed information at the end of the description''}). All co-authors then reviewed the original informal groupings, discussing and proposing new groupings to iteratively synthesize a set of design goals. Conducting an informal thematic analysis, we collaboratively sought goals such that each goal addressed a distinct user need and, together, the set maximally covered observations from the prior study. In the end, we identified the following four ways in which users would like to be able to customize narrated descriptions:

\begin{enumerate}
    \item \textbf{DG1: Presence}. Users described wanting to be able to choose which pieces of information were included in the narration.
    This is important because different types of information are useful for accomplishing different tasks\,---\,for example, one user may want the overall trend described in easy-to-understand terms, another may want statistical information like the average and standard deviation, and a third may want to skip those summaries and only hear about actual data points \cite{lundgard_accessible_2022}. Including all three types of information, with no option to turn off what is not wanted, could result in a narration so long that it would be overwhelming to some users \cite{jung_communicating_2021}. 

    \item \textbf{DG2: Verbosity}. Users also wanted control over the length and conciseness of the information that is present, generally preferring lower verbosity. Phrases with lower verbosity might use more abbreviated grammar or leave out common words (for example, the high verbosity phrase "the value for the 'x' field is 5" could have the lower-verbosity equivalent "x: 5"). For users who are less familiar with the conventions and forms of data visualizations, high verbosity is crucial for helping them understand what they're hearing. However, extra verbosity can slow users down significantly. This is due to the fact that screen readers narrate linearly, meaning that users cannot skip irrelevant content or skim back and forth \cite{cesal_writing_2020}.

    \item \textbf{DG3: Ordering}. 
    Users wanted to control the order in which information was presented, especially if they were looking for a specific piece of information.
    Similarly to verbosity, ordering is important because screen readers impose a linear reading order. Because screen reader users cannot skim through content, order has a stronger effect for them: if important information is left to the end, they must wait to hear it. This slows users down and increases the chance that they may miss information entirely, since they have to pay close attention to when the information being read out switches from being irrelevant to relevant.

    \item \textbf{DG4: Duration}. 
    We observed that users had different preferences about content depending on what they were looking for and how familiar they were with the representation's structure.
    Both of these factors can change over time, so the length of time that a customization lasts is an important consideration. A customization that helps a user with a particular task would ideally be easy to end when the user transitions to a different task, but a customization that accommodates a user's long-term preferences (for example, an expert hiding information about the shape of the structure) should last until the user turns it off, rather than requiring them to re-apply it every time they engage with a visualization.
\end{enumerate}

\subsubsection{Limitations}
Our design goals were derived from the perspective of a small group of researchers, of whom only one is blind. The population of 13 users we interviewed for our evaluation offered a wider variety of perspectives, and future work should take their opinions into account not only evaluatively but also during the initial phases of design. Our process was also an informal one; while we do not believe that this weakens our results, it’s possible that another approach with a different analysis method would offer an additional perspective on the matter.

\subsection{A Conceptual Model for Customization} \label{conceptual_model}

\definecolor{operator}{HTML}{008000}
\definecolor{codegray}{HTML}{545454}
\definecolor{symbol}{HTML}{007299}
\definecolor{literal}{HTML}{AA5D00}
\newenvironment{allintypewriter}{\ttfamily\color{codegray}}{\par}

\begin{figure}[h]
  \centering
  
  \begin{allintypewriter}
  \begin{tabular}{l}
\textcolor{symbol}{Customization} \textcolor{operator}{:=} (\textcolor{symbol}{Token}[], \textcolor{symbol}{Duration}) \\\\

\textcolor{symbol}{Token} \textcolor{operator}{:=} (\textcolor{symbol}{Affordance}, \textcolor{symbol}{Direction}, \textcolor{symbol}{Brevity}) \\

\textcolor{symbol}{Affordance} \textcolor{operator}{:=} \textcolor{symbol}{Wayfinding} | \textcolor{literal}{consuming} \\ 

\textcolor{symbol}{Wayfinding} \textcolor{operator}{:=} \textcolor{literal}{location} | \textcolor{literal}{surroundings} \\ 

\textcolor{symbol}{Direction} \textcolor{operator}{:=} \textcolor{literal}{upwards} | \textcolor{literal}{in-place} | \textcolor{literal}{downwards} \\ 

\textcolor{symbol}{Brevity} \textcolor{operator}{:=} \textcolor{literal}{low} | \textcolor{literal}{medium} | \textcolor{literal}{high} \\\\

\textcolor{symbol}{Duration} \textcolor{operator}{:=} \textcolor{literal}{persistent} | \textcolor{literal}{ephemeral}
  \end{tabular}
  \end{allintypewriter}
  \caption{Our design specification for customization. We define customizations that meet the four design goals we identify: they contain an ordered list of tokens, which addresses \textit{presence} and \textit{ordering} (DG1, DG3); each token has its own brevity, controlling \textit{verbosity} (DG2), and each customization has a \textit{duration} (DG4).}
  \label{specification}
\end{figure}

We now present a more formal definition of customization. 
We distilled this definition by referencing literature about user needs and tasks for data visualizations. We used Brehmer and Munzner’s typology of user tasks \cite{brehmer_multi-level_2013} as a basis for identifying different user needs and relating them to token affordances. We also referenced work specific to BLV users \cite{zong_rich_2022, lundgard_accessible_2022, jung_communicating_2021, morris_rich_2018}, since their needs and the affordances available to them may differ from sighted users. We reviewed the affordances of commonly-used screen readers like VoiceOver and NVDA, with the goal of developing customizations that were easy to use with typical screen reader software and whose design would be familiar to users. We iterated on a starting definition by using it to evaluate real-world examples, with the goal of finding categories which were both exhaustive (such that we could place every example within them) and meaningful (such that the groupings of examples highlighted real differences between groups). 

Our blind co-author and co-designer Hajas was instrumental during this iteration process, sharing his perspective as a researcher and a screen reader user. Our co-design process involved approximately biweekly regular discussions, both synchronous and asynchronous, between Hajas and the other co-authors over a period of six months. Before each discussion, the other co-authors would share a piece of our in-progress conceptual model and an implementation prototype with affordances matching that definition. Hajas would discuss the model with us, work through the prototype and offer his thoughts: where the definition failed to describe some important affordance and where it captured them well, how the prototype matched the definition or diverged from his expectations after reading it, and how we might consider refining both in order to better capture screen reader users’ needs and interaction preferences.

\subsubsection{Token}
We model a customization as a collection of tokens, where each token communicates a single piece of information. This approach is necessary for our goals of presence and ordering (DG1, DG3): if pieces of information are not recognized as distinct, a user has no way to choose which information they want to hear and in what order. To support a variety of information-seeking tasks, we define two parameters which determine the information each token conveys: the \textit{affordance} it provides the user and \textit{direction} in a hierarchy it provides the affordance for. To be able to support as many user tasks as possible, a system should provide a token for every combination of affordance and direction.

\textbf{Affordance.}
The affordance of a token corresponds to what type of lower-level task it allows a user to carry out. Drawing on Brehmer and Munzner's typology of user tasks~\cite{brehmer_multi-level_2013}, we divide tasks into two categories of affordance: \textit{wayfinding} and \textit{consuming}. \textit{Wayfinding} affordances are those that help the user find something they are looking for (whether or not they know the name or location of what they seek \cite{brehmer_multi-level_2013}). We further divide wayfinding affordances into two subcategories: \textit{location} and \textit{surroundings}. \textit{Location} affordances help the user answer the question ``Where am I?'' by providing information about what subset of the data they are currently viewing (for example, a token like ``values from 2004 to 2006''). \textit{Surroundings} affordances help the user answer the questions ``Where can I go?'' and ``What will I find?''; for example, a token like ``view 1 of 5'' answers the first question by giving information about the context outside the node; ``24 values'' answers the second question by giving ``information scent'' \cite{pirolli_information_1999} that helps the user know what to expect if they move downwards in the tree. \textit{Consuming} affordances are those which directly communicate data, for example ``price: \$400'' or ''the average temperature is 82 degrees Fahrenheit''.  Typically, when carrying out a high-level task like learning about a particular company's stock price, users will begin by using wayfinding affordances to find the area they're looking for, then use consuming affordances to get the information they want. Differentiating tokens by which part of this process they help with allows users to switch off tokens that aren't helpful or reorder the helpful tokens to be at the front, reducing unnecessary information and increasing efficiency.

\textbf{Direction.}
Although motivated by the hierarchical textual structures we described in \S\ref{sec:background}, we believe direction is central to customizing screen-reader-accessible visualizations as it enables an ``overview first, zoom and filter, and details on demand'' information-seeking behavior~\cite{shneiderman_eyes_2000}. 
Thus, the direction parameter affects the filters that are applied to data, changing what subset of it is selected and conveyed by a token.
The \textit{downwards} direction, then, corresponds to receiving information about filters that can be applied to the current data and the selections that would be produced. 
The \textit{upwards} direction corresponds to information about filters that were applied earlier in the hierarchy. 
The third direction, \textit{in place}, provides information about the current selection of data and the filter that produces it (as differentiated from the selections of sibling nodes in the hierarchy). 
Differentiating tokens by their direction allows users to toggle the \textit{presence} of which sets of filters they want to hear about (DG1). 
It also allows users to assign tokens a priority in the \textit{ordering} based on direction (DG3). For instance, information about upwards-facing filters can be important for establishing context, but can turn repetitive if every piece of content at the same level of abstraction continues to convey it.
Instead of toggling upwards tokens off, the user can simply push it further back in the ordering. 
That way, they're able to choose to hear that information if they need to be reminded, but can still easily skip it.

\textbf{Brevity.}
Brevity refers to the descriptiveness of content, and screen readers have built-in brevity options (sometimes called `verbosity') that typically toggle how much information is included. 
As summarized above, brevity is important so that users who need additional clarity can get it without slowing down users who don't need it (DG2). We instantiate brevity on the token level rather than the customization level. This granularity enables users to assign separate brevity settings to different combinations of affordance and direction. Consider two users who are both browsing through a graph with the general goal of learning about its trend. A more experienced user might set wayfinding tokens to a shorter brevity, since they are comfortable navigating, and set consuming tokens with summary statistics to a longer brevity, since they want to go deeper into the statistics. A less experienced user might configure longer wayfinding tokens and shorter consuming tokens because they're somewhat interested the statistics, but are more focused are finding their way around the graph.

\subsubsection{Duration}
We instantiate duration on the customization level, rather than the individual token level. This is because the duration of a customization should match the duration of the user's preference; we assume each customization is an expression of a single preference, so it should also have a single matching duration. This is also in line with user expectations. A single customization may have multiple effects (e.g. removing one token, changing another's brevity, and moving a third up to the front), but the user will execute it as a single action. Therefore, they will expect the results of their action to be atomic: it would be unexpected and confusing to them if one part of the customization ended and another one was still in effect. To fit with this expectation, every part of the customization should last for the same length of time and then expire simultaneously. 

We define duration to be either \textit{persistent} or \textit{ephemeral}. Persistent customizations should last indefinitely, until the user chooses to revert them; these are intended to support long-term user preferences which change over the scale of months or years (DG4). Ephemeral customizations may end or revert on their own; for example, if a user is applying a customization solely to the current node it may expire when they leave the node. At their longest, they last for the entirety of the user's current session but end when the user leaves the visualization. These are intended to support short-term user preferences corresponding to a single task, which likely lasts for only a few minutes.
\section{Implementing Customization in Olli}

We instantiate our conceptual model of customization as a set of extensions to Olli~\cite{blanco_olli_2022}, an open source toolkit for producing hierarchically structured textual descriptions of visualizations~\cite{zong_rich_2022}. Previously, the text at each node of the hierarchies generated by Olli was static. We redesign the text to be generated from a set of tokens and provide two ways to customize the properties of the tokens: a \textit{settings menu} for persistent changes and a \textit{command box} for ephemeral ones. This implementation was developed during the co-design process detailed in \autoref{conceptual_model}.

\begin{figure*}[b]
  \centering
  \includegraphics[width=\linewidth]{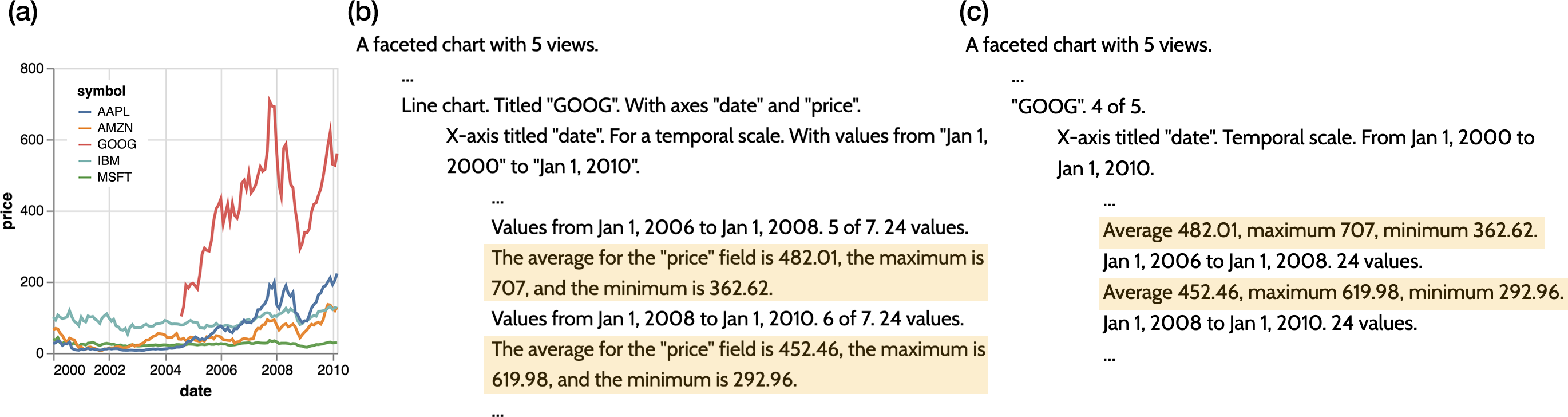}
  \caption{Two different customizations of Olli hierarchies for a chart showing five technology companies' stock prices between 2000 and 2010. (a) the visualization; (b) a customization that includes more tokens, with longer brevity, more suitable for novice users who need additional assistance in forming the correct mental model of the graph; (c) a customization with fewer, brief tokens more suitable for an expert user who might have a well-formed mental model of the chart.}
  \Description{A graphic with three parts labelled a, b, and c. Part a is a multi-series line chart showing stock prices of tech companies, with an x-axis titled date, and a y-axis titled price. Parts b and c each show part of an Olli hierarchy as indented lines of text. Each line of text is a node, and nodes lower in the hierarchy are further indented to show their depth. The nodes of the hierarchy in part b have more text, with sentences written out in full, and the nodes of the hierarchy in part c have less text, with more abbreviated sentences.}
  \label{stocks}
\end{figure*}

\subsection{Tokens}
In the non-customizable version of Olli, the text for each node is generated as a single block, and its content is dependent on the node's hierarchy level. In our extensions, the text for each node is generated from a list of tokens. This supports the \textit{presence} goal, as the user can control which tokens are included (DG1). The set of valid tokens for a node depends on its hierarchy level --- not every level can have every token, since the top and bottom levels have no downwards and upwards direction respectively. We defined a set of tokens covering all combinations of the \textit{affordance} and \textit{direction} parameters, and determined which tokens were applicable to each hierarchy level.

Table \ref{tokens} shows the set of implemented tokens. Each one is listed with its name and an example of what its text might be. For some cells in the table, we implement the only possible tokens: for example, the combination of the \textit{location} affordance and the \textit{in place} direction produces a token naming the values that define the current node's selected data. However, some cells have more than one possible token. In this case we drew on literature about user tasks as well as our iterative co-design process to choose the tokens that are most helpful for typical user tasks. For example, the combination of the \textit{consuming} affordance and the \textit{downwards} direction could provide many kinds of summary information to the user about the data of the selections available in the hierarchy below their current node. However, users typically want to hear about the overall trend of the data and its extrema \cite{sharif_understanding_2021} stated in plain language \cite{kim_accessible_2021}, and the average, minimum, and maximum of the selection satisfies these criteria. \autoref{stocks} shows an example of what these tokens might look like for two different use cases. 

\newcolumntype{Y}{>{\arraybackslash}X}

\begin{table*}
\begin{tabularx}{\textwidth}{|p{1.4cm}|p{1.8cm}|Y|Y|Y|}
\hline
& & \multicolumn{3}{c|}{\textbf{Affordances}} \\
\cline{3-5}
& & \multicolumn{2}{c|}{\textbf{Wayfinding}} & \multicolumn{1}{c|}{\textbf{Consuming}} \\
\cline{3-4}
& & \textbf{Location} & \textbf{Surroundings} & \\
\hline
\multirow{4}{*}[-2.5em]{\textbf{Direction}} & \textbf{Upwards} & Parent / Facet name (“MSFT”) & Depth ("Level 3") & Context (e.g. quantile, “3rd quartile”) \\
\cline{2-5}
& \multirow{2}{*}{\textbf{In Place}} & Name of current node (“X-axis”, “2000 to 2002”) & Index (“1 of 5”) & Data values (“price 65, date Jan 1 2010”, “range from 0 to 800”) \\
\cline{5-5}
& & & & Object type (“temporal scale”, “line chart”) \\
\cline{2-5}
& \textbf{Downwards} & Name of child nodes (“Axes date and price”) & Child size (“2 axes and 1 legend”, “10 intervals”) & Aggregate value (average, min, max) \\
\hline
\end{tabularx}
\caption{\label{content-tokens-table} A table demonstrating how the tokens we implemented in Olli provide coverage over possible combinations of affordance and direction. Each token is shown with its name and an example token. Where multiple tokens could fit in a cell, we chose tokens we predicted would be most helpful for users, with the goal of having the full set of tokens encompass the majority of information that most users need about a visualization.}
\label{tokens}
\end{table*}

\subsection{Settings Menu}

\begin{figure}[h]
  \centering
  \includegraphics[width=3.5in]{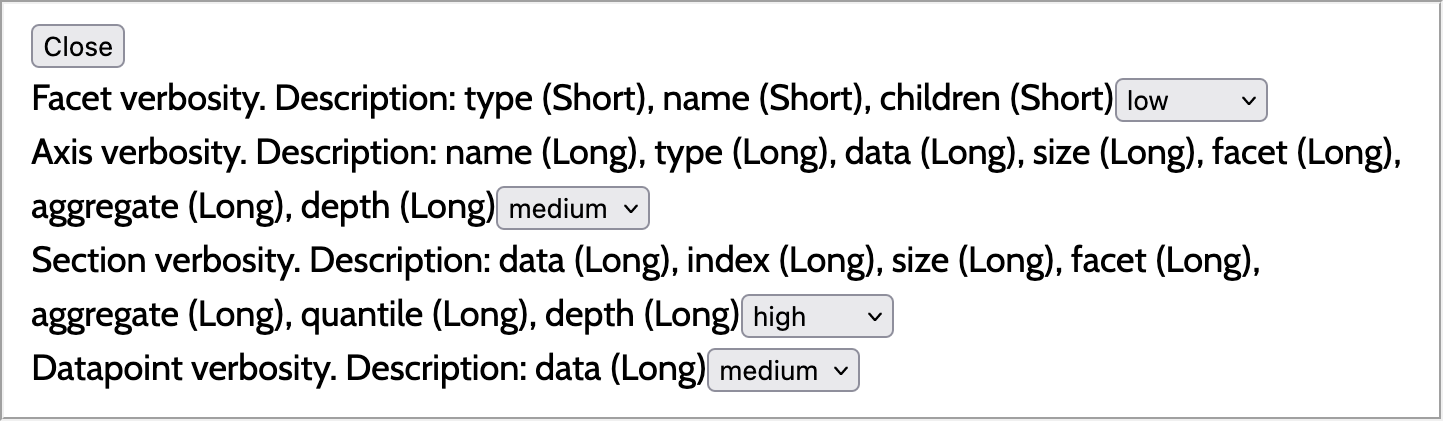}
  \caption{The Olli settings menu. ``Facet'', ``Axis'', ``Section'', and ``Datapoint'' correspond to the four levels of the Olli hierarchy. The user can set a separate persistent customization for each level, with three default options of high, medium, and low, as well as the option to create new customizations.}
  \Description{A screenshot of a settings menu. It has a close button at the top and then four lines of text. Each line has the name of a level in the hierarchy, then a description of the current customization for that level, and finally a dropdown menu for users to change the customization of the level. Each customization description is formatted as a list of tokens' names and their verbosity level, for example "depth (long)".}
  \label{settings}
\end{figure}

The settings menu in \autoref{settings} allows the user to make persistent customizations to the presence (DG1), brevity (DG2), and ordering (DG3) of tokens. We use a settings menu because screen readers offer similar menus for their own customizations, making it a familiar interface for users. The settings menu has four sections, each of which corresponds to one of the Olli hierarchy levels (``Facet'', ``Axis'', ``Section'', and ``Datapoint'' in \autoref{settings}). For each hierarchy level, the user can use a dropdown menu to choose between one of three preset customizations that we define, named \texttt{high}, \texttt{medium}, and \texttt{low}. These customizations are intended to meet the needs of relatively novice users who are performing common tasks, typically trying to get an overview of the trend in data and then zooming in on individual points in one or two areas. For example, they may use the \texttt{low} or \texttt{medium} setting to form a model of the shape of the graph, turn to \texttt{high} to hear more about the trend and summary statistics, and turn back to \texttt{medium} to look at individual data points.

So that the user can understand the differences between customizations, a label for each dropdown menu describes the currently active customization. The description states which tokens are included and whether their individual brevities are \textit{short} or \textit{long}. The \texttt{high} customization includes all possible \textit{long} tokens for a given hierarchy level. The \texttt{medium} customization includes only a subset of tokens, all set to \textit{long} brevity; the \texttt{low} customization includes the same subset, but all set to \textit{short} brevity.

\begin{figure}[h]
  \centering
  \includegraphics[width=\linewidth]{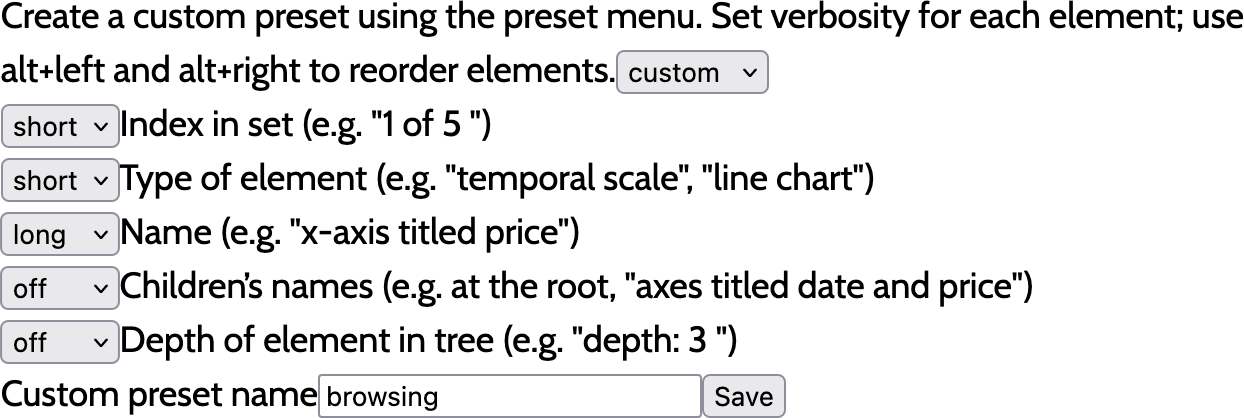}
  \caption{The user interface for creating a new customization for the settings menu. Each customization is specific to one hierarchy level. For each token that can be included in that hierarchy level, the user can choose whether to exclude it, include it with a short brevity, or with a long brevity. They can also choose to reorder tokens. This creates a complete customization that meets the \textit{presence}, \textit{verbosity}, and \textit{ordering} goals (DG1, 2, 3).}
  \Description{A screenshot of the interface for creating a custom preset. It shows a set of five dropdowns. Each dropdown controls the customization for one particular token and has the options off, short, and long. Next to each dropdown is a description showing the name of the corresponding token and giving an example of that token. At the bottom is a text box labelled "custom preset name" and a Save button. The current user has filled in "browsing" as their preset name.}
  \label{custom}
\end{figure}

To further support users with more specialized tasks, we offer a more in-depth customization option: in addition to the built-in \texttt{high}, \texttt{medium}, and \texttt{low} customizations, a user can create their own customizations. Each customization applies only to a single hierarchy level, and each hierarchy level can have unlimited customizations. The interface for creating a customization is pictured in \autoref{custom}. It contains one dropdown menu for each token that can be included in the hierarchy level. Each dropdown menu has the settings \textit{off}, \textit{short}, and \textit{long}. Tokens that are set to \textit{off} are not present in the customization; \textit{short} and \textit{long} tokens are present at the corresponding brevity (DG2). The user can choose the order of the tokens by reordering the menus with a keyboard shortcut (DG3). After they assign a name to the customization and save it, it will appear in the list of possible settings for that hierarchy level and act as a persistent customization, with no difference between user-created customizations and the three built-in options. 

These custom presets are intended for more experienced users and allow them to accomplish two things at different levels of duration (DG4). First, they can set up presets that match their overall preferences for navigable hierarchies. For example, if a user's screen reader tells them their current depth in the hierarchy already, as some do, the user can turn off the depth token in all of the customizations they create. Second, the user can set up customizations that match particular specialized tasks. For example, if a user works with data visualizations for their job and carries out a few primary tasks on these visualizations, they could have a customization for each task and switch between them depending on the work they're doing that day.

\textbf{Design Considerations.}
The settings menu is designed as a ``bounded space'' \cite{zong_rich_2022}: it is opened and closed using keyboard commands or buttons, but the user can't shift their focus to access the menu when it is closed or leave it when it is open. This means that when the menu is closed, it doesn't get in the user's way as they try to explore the main purpose of the page and the visualization. When it is open, the user can't accidentally shift their focus outside of the menu, leaving them unsure of whether their changes have been successfully applied. This aligns with prior findings that bounded spaces ``alleviate cognitive load by allowing a user to maintain their position relative to entry points'' \cite{zong_rich_2022}.

\subsection{Command Box}
\begin{figure*}[h]
  \centering
  \includegraphics[width=\textwidth]{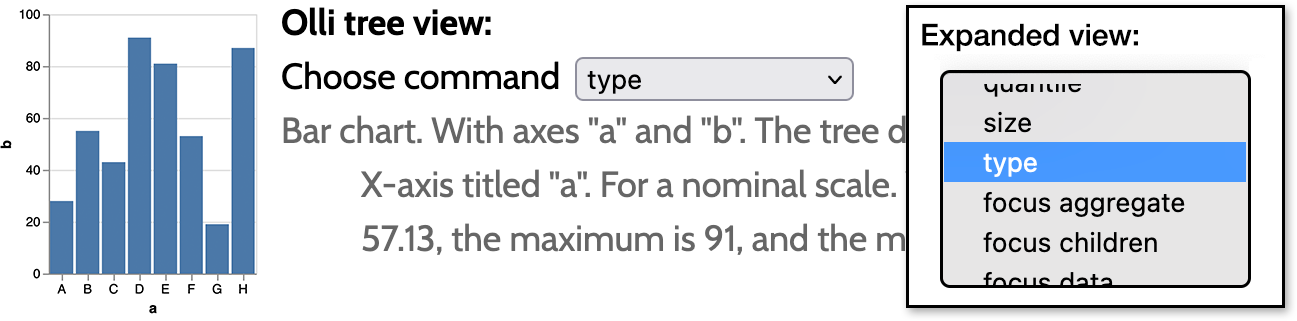}
  \caption{The Olli command box in use for a visualization of a bar chart. The box is expanded to show the end of the set of \textit{speak} commands and the beginning of the set of \textit{focus} commands. The user can use their arrow keys to select a command, or type the beginning of the word, and then hit enter to execute the selected command. This quick execution method supports the \textit{duration} goal by providing a way to do ephemeral customizations without significant overhead.}
  \Description{A figure with three parts showing the Olli command box. The first part is a standard visualization of a bar chart without any particular type of data specified. The middle part shows an Olli hierarchy with a dropdown in focus. The dropdown has the label "choose command", and the selected option is "type". The third part shows what the dropdown looks like when it is open and showing multiple options. Four options for commands are visible: "size", "type", "focus aggregate", and "focus children". Other options are cut off at the bottom and top of the dropdown, showing that more commands are available if the user scrolls up or down.}
  \label{command-box}
\end{figure*}

The command box allows the user to make ephemeral customizations to the presence, brevity, and ordering of tokens. Whereas the goal of the settings menu is to support users in making in-depth customizations to their long-term experience of data visualizations, the goal of the command box is to support users in switching tasks quickly as they move through a single visualization task (DG4). Therefore, it is implemented as one dropdown rather than a series of dropdowns as in the settings menu. The user can choose a command from the dropdown and hit enter to execute it.

There are three types of command: \textit{speak token}, \textit{focus token}, and \textit{shortcut settings}. \textit{Speak} commands do not change the text in the hierarchy itself, but instead speak out the value of the chosen token for the current node in the hierarchy. The effect of these commands end as soon as the token is done being read aloud. Speak commands support the \textit{presence} goal (DG1): users may not want to include a token for the entire hierarchy, but need it for one particular node to learn something specific. These commands also have a practical function: if the user missed one particular piece of information when the node was being read aloud, they can hear it again without having to replay the entire text. \textit{Focus} commands rearrange the text in the hierarchy, so that the previous ordering is maintained except that the chosen token is now first. Multiple focus commands can stack, and the effect of these commands is active either until the user leaves the visualization, or until they use the \textit{clear} command to clear all focuses. Focus commands support the \textit{ordering} goal (DG3) by allowing the user to quickly change the order of one or a few tokens that are important to their current task. This lets the user quickly move between lower-level tasks without needing to fully design and switch between custom settings. Finally, \textit{shortcut} commands are simply shortcuts to the settings menu. The user can use these commands to apply a customization to a hierarchy level (whether \texttt{high}, \texttt{medium}, \texttt{low}, or a custom setting). These are available for convenience; because they are actually settings menu customizations, their effects are persistent.

\textbf{Design considerations.} 
Throughout our design for both the settings menu and the command box, we found ourselves using dropdowns in cases where sliders (e.g. off/short/long when creating customizations) or text boxes (e.g. for entering commands) might commonly be used. Dropdowns have the benefit of being both discoverable and efficient. They are discoverable because they present all of the available options to the user in a list that the user can easily move through using the arrow keys. Some sliders may support this functionality, but not all do, and text boxes typically do not. Discoverability is particularly helpful for novice users who don't yet know what options are available to them. Dropdowns are also efficient, because they provide built-in autocomplete functionality: users can start typing the word they have in mind and the selection will jump to the first option in the dropdown that starts with those letters. Sliders do not typically have this option, and implementing it on text boxes requires additional effort or specialized third-party libraries not built with accessibility in mind. Autocomplete is particularly helpful for expert users who know what they want and want to move to that option as quickly as possible, and in cases like our command box where the number of options is high enough that moving through all of them by hand is time-consuming.

\section{Evaluation}
To evaluate our contribution, we recruited 13 blind and low-vision users and conducted a 90-minute Zoom interview with each participant. Participants were asked to explore Olli's settings menu and command box, each on a different dataset, and perform tasks using them. The goal of our evaluation was to determine whether the customization affordances provided improved users' ability to complete tasks efficiently and to observe how the user's preferences and task at hand influenced their customization choices.

\subsection{Study Design}
Jones, Pedraza Pineros, and Zong conducted the interviews using Frøkjær and Hornbæk’s Cooperative Usability Testing (CUT) method \cite{frokjaer_cooperative_2005}: Jones acted as the guide, talking to the user and explaining the prototypes, and Pedraza Pineros and Zong acted as loggers noting usability problems and relevant statements from participants. 

\textbf{Study Setup}. Each interview was 90 minutes. We began by introducing the participant to Olli's navigable text hierarchy without any customization, to familiarize them with its interface.
The bulk of the session was spent on evaluating the settings menu and command box interfaces in turn\,---\,following the same protocol for each condition. First, the guide showed the participant a multi-series line chart and walked the participant through using the interface on it for the first time. Next, we gave the participant time to do open-ended exploration of the interface by prompting them to ``explore the tool and try it out.'' We then asked participants to complete two specific tasks on the data and watched how they approached the tasks. After finishing this process for both interfaces, we conducted a semi-structured discussion asking the participant about their use of customization in their daily lives.

For each interface condition, participants were set two tasks: find which series in the dataset contained the extrema, and summarize a trend for a single series. 
We chose these tasks to fully exercise our two interfaces by forcing participants to move through multiple smaller tasks spanning much of Brehmer and Munzer’s task typology \cite{brehmer_multi-level_2013}. 
The two tasks are complementary in that the first requires participants to find one datapoint across many series, and the second many datapoints across one series. 
Both tasks allow for multiple ways to approach the data to find the same answer\,---\,for example, getting a sense of data values over time for a series could be done by reading summary statistics for one or a couple nodes at a higher level in the hierarchy, or by quickly flipping through a dozen individual data points at a lower level. 
Our hope was that picking two very different tasks, each with more than one possible approach, would let us see how our interfaces performed in different situations and with users who preferred different problem-solving techniques. 

\textbf{Participants} We recruited 13 blind and visually impaired participants through our collaborators in local blind community organizations and through a public call on Twitter. Each participant received \$50 for a 90-minute Zoom session. Here we provide aggregate participant information to protect privacy \cite{saunders_anonymising_2015}, and with the acknowledgement that socially-constructed data like race/ethnicity should be collected and publicized with care \cite{chen_why_2023}. 77\% (n=10) of participants self-identified as totally blind, 15\% (n=2) self-identified as totally blind with some light perception or low vision, and 8\% (n=1) did not respond. 69\% (n=9) have been blind since birth. Participants were split into 31\% (n=4) NVDA users, 62\% (n=8) JAWS users, and 8\% (n=1) Voiceover users, aligning with screen reader statistics \cite{noauthor_webaim_2021}. In addition, 77\% (n=10) used Google Chrome as their browser, and 23\% (n=3) used Firefox as their browser. Demographically, 39\% of our participants use she/her pronouns (n=5) and the rest used he/him pronouns (n=8). Participants self-reported their ethnicities (Asian, Black/African, Hispanic/Latinx, and  Caucasian/white), covered a diverse range of ages (20--50+), and had a variety of educational backgrounds (undergraduate, graduate, and trade school). 10 participants self-reported as slightly to moderately familiar with statistical concepts and three as expertly familiar. Ten participants self-reported as slightly to moderately familiar with data visualization concepts and methods, one as not at all familiar, and one as expertly familiar. Participants reported a high variety of frequency interacting with data or visualizations, from never to 3 or more times/week, with most reporting 1–2 times/month. Most participants (n=7) reported that they rarely use data analysis tools, but some (n=2) considered them an important part of their workflow. 

\subsection{Quantitative Results}

\subsubsection{Likert Scales}

\begin{table*}
  \caption{Rating scores for each prototype (Settings Menu, Command Box) on a five point Likert scale where 1 = Very Difficult (Very Unenjoyable) and 5 = Very Easy (Very Enjoyable). Median scores are shown in boldface, averages in brackets, standard deviations in parentheses.}
  \label{tab:likerts}
  \begin{tabular}{lcc}
    Prompt: After understanding how the [prototype] works... & Settings Menu & Command Box \\
    \midrule
    How easy was it to learn to customize? & \textbf{4} [3.69] (0.75) & \textbf{4} [3.62] (1.04) \\    
    How enjoyable was it to interact with the data? & \textbf{4} [3.54] (0.78) & \textbf{4} [3.62] (1.26) \\ 
    How easy was it to find the place you were looking for in the description? & \textbf{4} [3.46] (1.05) & \textbf{4} [3.54] (1.27) \\ 
    How easy was it to get the information you wanted? & \textbf{3} [3.31] (0.95) & \textbf{4} [3.62] (1.19) \\ 
    How easy was it to understand trends and patterns in the data? & \textbf{3} [3.31] (1.11) & \textbf{4} [3.46] (1.27) \\ 
    \bottomrule
  \end{tabular}
\end{table*}

To evaluate how well each prototype met our four goals for customization, we designed a Likert survey to understand a participant’s ease of \textit{wayfinding} and \textit{consuming} data. Participants responded on a five point scale where 1 = Very Difficult/Unenjoyable and 5 = Very Easy/Enjoyable (\autoref{tab:likerts}).
The median scores suggest that participants generally found customization, wayfinding, and consuming easy for both prototypes. When it came to more complex tasks, i.e. interpreting the data, participants found the command box easier to use than the settings menu. Notably, the settings menu prototype was shown before the command box and before the participant’s 5-minute break; by the time participants began using the command box, they could have had a greater familiarity with the Olli tree view, explaining why the command box, in general, had better reviews. We contextualize the reasons behind participants’ scores through qualitative analysis.

\subsubsection{Action Logging}

To understand how participants used different customization options, we collected a log of participants' interactions with each prototype. For the settings menu prototype, 37\% of customizations applied across all hierarchy levels used the \texttt{low} preset, 37\% used \texttt{medium}, and 16\% used \texttt{high}. 11\% of the time, users applied a user-defined custom setting. This supports our findings in \ref{qual-results-characteristics}, as users preferred to minimize brevity when possible. 

For the command box prototype, we broke down participants' logs into three major categories: \textit{presence}, \textit{order}, and \textit{brevity}. 69\% of users' commands toggled \textit{presence}, 22\% changed \textit{ordering}, and 10\% adjusted \textit{brevity}. This aligns with the idea that users preferred to use the command box's ephemeral customizations to adjust presence and order on the fly. On the other hand, brevity is likely more frequently set using the settings menu's persistent customizations.

\subsubsection{Limitations}
As participants filled out the Likert scale surveys, they often narrated their thought processes in ways that helped us surface limitations of our quantitative study results.
For instance, the limited amount of time they had to both become familiar with Olli navigation, and to learn to customize, meant that their scores represent only a snapshot of their learning process. This has effects in both the positive and negative direction. For example, while some participants who had trouble with Olli gave lower scores  to express their difficulty navigating, others gave higher scores when they found something difficult but believed they would derive value from it once they had more time to learn.

\subsection{Qualitative Results}

After finishing interviews, Jones, Pedraza Pineros, and Zong individually performed open coding, using a grounded theory approach \cite{charmaz_constructing_2006}, on the notes taken by the logger in each interview. All three then met to review the codes. The number of codes identified by each author were 11, 13, and 14; of these, 5 codes (26\% of the unique codes) were identified by all three authors, 9 (47\%) were identified by two authors, and the remaining 5 were only identified by one author. We took all themes identified by two or more authors and grouped them into 3 higher-level groups\,---\,or example, \textit{``learnability is a barrier''} and \textit{``different help for different people''} were both grouped into \textit{``the usefulness of customization is context-dependent''}. These groups and the themes that made them up formed our final qualitative themes.

\subsubsection{Customization supports autonomy and agency, yet opportunities to customize are often lacking.}\

\textbf{Customization supports autonomy and agency.} Blind and low vision users, more so than sighted users, often feel ``at the mercy'' (P7) of those who design the systems that they use. P7 remarked that designers ``just assume what we want and how we want it laid out'', and P12 that ``alt text writers make assumptions about what data I will be interested in.'' Customization lets them get the information they need and can be the difference between being out of the loop or being able to react to important news --- for example, participants who wanted not only to hear overall COVID statistics but also to be able to learn about the statistics for their area in particular. It can be the difference between being able to do a part of one's job or having to rely on others to do so. P9, discussing her recent move to a more data-heavy position, mused ``Am I going to depend on collaborators to do all the analysis [...] or am I gonna keep trying to find ways?'' This decision was a ``work in progress'' for her, in part because of the inaccessibility of the data analysis tools her position typically uses, which she contrasted with ``this kind of tool'' (our prototype) that might allow her to go deeper into the data.

\textbf{However, most screen reader interfaces lack domain-specific customization.} Screen readers come with built-in customization options which all participants said they used at least occasionally. These options are typically customizing which punctuation are read out (for example, skipping parentheses) and what and how structural elements of a page, like headers and links, are read. However, these options typically apply over the user's entire computer or an entire app, despite the fact that users can use their computer for very different tasks: reading the news, email, research, etc. (These could all be within a single browsing app!) This forces users to manually switch between settings when switching tasks (P5 mentioned that he keeps both NVDA and JAWS on his computer so that he can have two sets of settings and more easily switch between them). P11 noted that he turns punctuation on when coding, since brackets are semantically important for code, but turns it off for song lyrics because ``I don't want to hear `hello comma is it me you're looking for'''. 

Moreover, screen reader customization options are not designed with affordances for particular domains of user task, even very common ones. P5 related that when listening to his email, he wanted to hear the subject read before the sender because if the subject didn't interest him he would skip the rest. However, the ability to re-order these elements was not available in JAWS. Domain-specific customization, like understanding that an email subject and sender are separate pieces of information that should be able to be controlled separately, are important to help users exercise agency in all areas of technology. We address this for data visualizations by allowing users to separately customize different levels of textual hierarchies and by dividing text into tokens based on users' tasks and needs (the data visualization equivalent of splitting up a subject and a sender).

\subsubsection{The four goals for customization that we identified are each important to users}
\label{qual-results-characteristics}\

\textbf{Presence.} Customization of \textit{presence} lets users choose which tokens they want to hear. Participants often expressed preferences about the presence of tokens even before being informed by the interview guide that they could control this, saying things like ``I don't need tree depth'' (P2, P12) or ``I love that it gives all the averages'' (P7). What task they were doing or planned to do had an impact on users' choice of tokens. P3 imagined potential tasks while choosing which tokens to include, but after being introduced to and completing the task given, said that ``having done this, I would go back and tweak the settings.'' Multiple participants adjusted their settings mid-task in order to hear tokens relevant to the task. In P3's case and others, this was because they'd begun a task by turning off summary tokens to reduce the length of the text and make wayfinding easier. Once they had found their desired area, they needed to turn back on the summary in order to efficiently consume the data and identify its trend. Some participants were able to do this easily, but others who were less confident in their customization struggled to switch affordances and found it more difficult to complete the task.

\textbf{Verbosity.} Discussing \textit{verbosity}, participants were perhaps more united than on any other point: as low a verbosity as possible is best. P4 explained it as ``the less speech you can get by on the better'', saying that ``if you can get the data you want without all that verbosity, it's just better. It's faster and less fatiguing.'' P7 further explained that this is particularly true for screen reader users: ``because you're listening, you're doing a lot more remembering. The ten seconds less that you spend [with a lower verbosity] is the difference between understanding and not understanding.'' However, sometimes the additional clarity of higher verbosity is needed. P9 specified that she prefers to switch to low verbosity only after she ``[gets] the gist'' of a data visualization and ``know[s] how the thing is laid out.'' P7 noted that ``if you're somebody who needs more contextual information, you can get that with a higher verbosity level.'' Participants would sometimes set verbosity levels to low, but become confused when trying to navigate the visualization because they weren't able to fully understand what each level of the hierarchy represented. After trying for a little while, they would set the verbosity levels back to medium and typically then have more success in understanding the hierarchy.

\textbf{Ordering.} Similarly to verbosity, \textit{ordering} is useful to users primarily because it improves efficiency and reduces the effort it takes to find needed information. P12 explained her interest in bringing the most important information up front: sighted people can ``see the far right of each line,'' but for her and other screen reader users, ``you've got to sit and listen to the screen reader before you get there.'' When verbosity was high and each node had a lot of text, participants would sometimes miss information towards the end of the node. After using the command box to bring information to the front, P5 said ``I didn’t hear those averages before'' because they were too far towards the back. He had also not needed the averages while forming his mental model of the graph; it was only when asked to complete a task involving finding a trend that he went looking for a way to hear summary statistics. 
In general, participants' preferred ordering was task-dependent; for example, when trying to locate Chicago, P2 appreciated that the command box ``allowed [her] to put city first.''

\textbf{Duration.}
Participants agreed that the settings menu, while useful, required too much effort to use frequently. P7 made a setting only for the axis section of the hierarchy because ``I spend more time in the axis,'' and P13 agreed that she ``do[es] customize things when [she uses] them a lot.'' P1 and P8 agreed that they ``wouldn't want to go back to do this again too often'' (P1). P11 explained that his decision about when to use the settings menu is ``task based'': ``I set it and then forget it for a while until I need to do something that requires some other setting.'' In contrast, participants appreciated the quick and on-demand nature of the command box, saying that ``I like that I can do it on the fly'' (P7) and that the command box is helpful because it's ``much faster'' than the ``multi-layered task'' of moving through the settings menu (P9).

\textbf{Affordance.} When asked to complete higher-level tasks, participants typically required first wayfinding and then consuming affordances. When asked to discuss which parts of the task were easier or harder for them, they frequently split the work out in ways that match these affordances, for example saying that ``I understood the trends and patterns, but the problem was getting to the information'' (P2). P2 was successfully able to consume and understand the data, but was struggling with wayfinding. Having tokens that are divided along lines similarly to how users naturally divide up their tasks makes it easier for users to turn on only the tokens that match what they need. One particular example of this is that wayfinding acts as a pre-requisite for consuming, since users must ``[get] to the information'' (P2) before they can read it. The ability to reorder tokens is especially important for wayfinding since participants would often move around quickly, hearing only the beginnings of nodes, to get a sense of the entire graph before spending more time in any one area. Participants noted that they were much more successful at wayfinding after moving the correct token to the front.

\subsubsection{The usefulness of customization is context-dependent.}\

\textbf{Customization interfaces become more useful with time and experience.} It stands to reason that customization is easier and more helpful when users have a good understanding of both the interface being customized and the interface that does the customization. Many of our participants said that our tools would become more helpful and enjoyable once they had spent more time using them. P1 felt he would ``need a lot of time to play around,'' while P2 said that if she ``had like 5 more minutes,'' it would have been easier. P4 said that although it ``took [him] a second to figure out a strategy,'' he was able to do so and ``it'll eventually be fun to play with stuff like [our customization tools].'' P10, whose job involves data analysis, said that he views our tool as ``a product I would sit down and learn to use'' because it would likely ``be very worth my while professionally.'' 
Besides underscoring the importance of giving users time to explore tools and making features discoverable, this also has implications for study design. While our sessions were limited to 90 minutes, future studies may want to consider longitudinal designs where users have a chance to return to prototypes over multiple days.

\textbf{Users have different preferences about customizing and relying on defaults.} Our participants had very widely varying preferences about customization, ranging from P4 who ``tend[s] to get by on what [he has] and figure out ways not to worry about it'' to P10 who wondered about the possibility of giving the user their own ``scripting capabilities'' to produce additional tokens with more sophisticated statistical analyses. Most participants fell somewhere between these two extremes, typically expressing that in some situations default values worked well for them, but in others they preferred to customize. Which situations fall into which category varied. P3 said she waits to customize until she finds that defaults ``are missing a component I really feel I need,'' a sentiment shared by P8 and P9. P2 noted she often felt such a component to be lacking and therefore frequently turned to customization, and P5 and P7 agreed that they customize as much as possible. 

\textbf{The topic at hand affects users' interest in customization.} As Peck, Ayuso, and El-Etr find, ``data is personal'' \cite{peck_data_2019}: participants cared far more about customizing for situations or visualizations covering topics they cared about. P4, who almost never customized, said that he did so in one particular case: when his screenreader was mispronouncing the name of one of his favorite sports teams. During the course of the interview, users saw two different datasets, and many expressed that they were more likely to use customization to help them understand the data when it was a subject they were interested in. When looking at a graph of stock prices, P12 said ``as someone who doesn't know the stock market well, I don't know how I would customize here''. On the other hand, discussing her use of commands to get more information about a temperature graph, P9 noted, ``I'm a wicked weather nerd, so I'm actually interested in the numbers.'' P6 explained that his interest in customization ``would change depending on [his] familiarity with the table,'' and P13 agreed that her interest in customization ``depends whether what I’m doing is for a project or if I’m just passing something on the way.''

\textbf{Prior experience affects users' comfort with customization.} Because both data visualizations and customization are frequently inaccessible to BLV users, many of our participants lacked prior experience with the kinds of tools we asked them to use. When asked to compare their experience with our prototype to other systems they use to customize or to access data visualizations, participants were often unable to: P11 said that ``99\% of the time there's not a chance to customize anything,'' and P4 said that he usually ``ignore[s] charts and graphs'' rather than ``battle'' with ones that are inaccessible. A lack of experience can naturally lead to a lack of understanding, which incentivizes users to avoid customization (thereby perpetuating their lack of experience) rather than accidentally change a setting that will have negative consequences for their interface. P9 admitted that ``I'm very afraid of [creating] custom [configurations]'', and many participants preferred to stick to our presets (rather than creating their own customizations) because they required less familiarity with the interface and provided less opportunity to make a mistake. In contrast, participants who were more comfortable with data visualizations and customization were the most likely to express a greater frequency of and desire for customization. P7 called himself a ``power user'' and noted that ``as someone that has written software, someone that programs, someone that teaches,'' he relies on customization to get the information he needs immediately rather than being slowed down unnecessarily.

\subsubsection{Limitations}
Our prototype was certainly not perfect and we observed limitations with it during the course of user interviews. In our implementation, each customization applies to a single hierarchy level, with no way to change the settings for multiple levels with one action. This was intended to give users more fine-grained control, since some tasks might be easier with customization for specific levels.
Experienced users appreciated this fine-grained control. However, we observed that less experienced participants who were carrying out more general overview tasks did not: they tended to apply the same customization to all four hierarchy levels, one after the other, every time they changed their settings. This implies that a way to globally change the settings for the entire tree would have been helpful for these users. 
Another major limitation was that the command box had so many commands that participants struggled to remember all of the affordances it offered, sometimes expressing that they wished they could do something that they didn't realize actually \textit{was} possible.
Future work may want to investigate different methods for surfacing help and reminders in-situ while users are carrying out tasks.
\section{Conclusion and Future Work}

In this paper, we presented four design goals for how BLV people should be able to customize screen-reader-accessible visualizations. 
We distilled these design goals through a co-design process with our blind co-author, and through an analysis of the self-reported needs of study participants navigating accessible data representations. 
We built a model of how customization can be implemented to meet these four goals, treating text as a sequence of individual tokens which each have their own properties and can be customized individually. We instantiated this model of customization in Olli~\cite{blanco_olli_2022} and found that it improved our participants’ ability to access the information they sought. 
Our participants were, on the whole, delighted to be able to customize in accordance with our four goals. 
Participants sometimes even surfaced, unprompted and without knowing the goals we’d identified, that our implementation helped them meet those same goals. However, the biggest barrier for participants was the effort of learning to use our tool. This work was not evenly distributed: it was more difficult for those who did not have prior experience with navigable hierarchies or customizable accessibility tools. 

\subsection{Interpreting as a Distinct Token Affordance}
In the course of our conceptual work, we theorized the existence of a third affordance in addition to \textit{wayfinding} and \textit{consuming}: \textit{interpreting}. Tokens that have the interpreting affordance would allow users to take the data they consume and situate it within a broader context outside of the visualization (for example, the graph of stock prices could include the information that the dip in stock market prices during 2008 was linked to the U.S. recession during that time). A user already familiar with the data might be apply to supply this using their own knowledge, but less familiar users may miss these connections without explicit help. This affordance is built on Lundgard and Satyanarayan's four-level model of semantic content \cite{lundgard_accessible_2022}: lower levels of content are more grounded in the properties of the visualization and higher levels require more synthesis of the data. Wayfinding and consuming affordances cover the first three levels; we name \textit{interpreting} as covering the fourth and highest level, that of contextual and domain-specific knowledge. 
As Lundgard and Satyanarayan note, at present, this type of content is very difficult to generate automatically. For this reason, we were not able to develop a model and implementation including the interpreting affordance. 
However, recent developments in large language models have made accurate machine generation of higher-level semantic text significantly more possible, and in the near future we hope to explore the addition of this affordance to our model of customization.

\subsection{Multisensory Data Representations and Broader Implications of Our Conceptual Model}

Many users in our study reported using non-text modalities --- such as sonification, braille, and tactile graphics --- alongside their existing screen reader workflows. These modalities have sense-specific affordances and limitations, just as text does. Sonification, for instance, excels at providing an overview of data. It helps users ``get the whole picture quickly without having to arrow through every value'' (P9). However, it's less suitable for communicating specific values --- most users are not able to ``hear a sound and think, oh, it must be this color'' (P10).
When it comes to tactile representations, participants praised the ability to ``jump all around'' 
(P6), skim or read at their own pace (P4, P12), and help those with hearing loss (P7). However, they noted that tactile displays are ill-suited for charts with occlusion (e.g. lines that cross or overlapping scatterplot points) (P9), and that the high cost of braille displays (P6, P10) and labor to produce tactile graphics (P9) pose a significant barrier.
As these participants' comments show, users' tasks and preferences shape how they use data representations broadly, beyond just visualization and text.

Our interviews surfaced areas where a systematic approach to customization could potentially benefit users of other multi-sensory representations.
For instance, P10 said that spatial audio was important to their coding workflow, but P7 noted that spatial audio is not intelligible for people who have hearing loss in one ear. Inflexible sonification interfaces can also cause additional barriers. For example, P10 has synesthesia and associates sounds with specific colors. They lamented that in a sonification tool that maps color to sound, ``the sound that represents green sounds red to me,'' and they were unable to customize the mapping (P10).
Future work could explore how designers might instantiate our conceptual model for other senses, and how the affordances of tactile graphics or sonifications shape users' preferred customizations.

\begin{acks}

We thank Jerry Berrier of the VIBUG group for his help in recruiting evaluation participants, and Frank Elavsky for his feedback throughout our writing process. This research is supported by NSF Award
\#1942659.
\end{acks}

\bibliographystyle{ACM-Reference-Format}
\bibliography{access-description}


\begin{thebibliography}{34}


\ifx \showCODEN    \undefined \def \showCODEN     #1{\unskip}     \fi
\ifx \showDOI      \undefined \def \showDOI       #1{#1}\fi
\ifx \showISBNx    \undefined \def \showISBNx     #1{\unskip}     \fi
\ifx \showISBNxiii \undefined \def \showISBNxiii  #1{\unskip}     \fi
\ifx \showISSN     \undefined \def \showISSN      #1{\unskip}     \fi
\ifx \showLCCN     \undefined \def \showLCCN      #1{\unskip}     \fi
\ifx \shownote     \undefined \def \shownote      #1{#1}          \fi
\ifx \showarticletitle \undefined \def \showarticletitle #1{#1}   \fi
\ifx \showURL      \undefined \def \showURL       {\relax}        \fi
\providecommand\bibfield[2]{#2}
\providecommand\bibinfo[2]{#2}
\providecommand\natexlab[1]{#1}
\providecommand\showeprint[2][]{arXiv:#2}

\bibitem[noa({[n.\,d.]})]%
        {noauthor_screen_nodate}
 \bibinfo{year}{[n.\,d.]}\natexlab{}.
\newblock \bibinfo{title}{Screen {Readers}}.
\newblock
\newblock
\urldef\tempurl%
\url{https://www.afb.org/blindness-and-low-vision/using-technology/assistive-technology-products/screen-readers}
\showURL{%
\tempurl}


\bibitem[noa(2018)]%
        {noauthor_webaim_2018}
 \bibinfo{year}{2018}\natexlab{}.
\newblock \bibinfo{title}{{WebAIM}: {Survey} of {Users} with {Low} {Vision} \#2 {Results}}.
\newblock
\newblock
\urldef\tempurl%
\url{https://webaim.org/projects/lowvisionsurvey2/}
\showURL{%
\tempurl}


\bibitem[noa(2019)]%
        {noauthor_blindness_2019}
 \bibinfo{year}{2019}\natexlab{}.
\newblock \bibinfo{title}{Blindness {Statistics} {\textbar} {National} {Federation} of the {Blind}}.
\newblock
\newblock
\urldef\tempurl%
\url{https://nfb.org/resources/blindness-statistics}
\showURL{%
\tempurl}


\bibitem[noa(2021)]%
        {noauthor_webaim_2021}
 \bibinfo{year}{2021}\natexlab{}.
\newblock \bibinfo{title}{{WebAIM}: {Screen} {Reader} {User} {Survey} \#9 {Results}}.
\newblock
\newblock
\urldef\tempurl%
\url{https://webaim.org/projects/screenreadersurvey9/}
\showURL{%
\tempurl}


\bibitem[Ahmed et~al\mbox{.}(2012)]%
        {ahmed_accessible_2012}
\bibfield{author}{\bibinfo{person}{Faisal Ahmed}, \bibinfo{person}{Yevgen Borodin}, \bibinfo{person}{Andrii Soviak}, \bibinfo{person}{Muhammad Islam}, \bibinfo{person}{I.V. Ramakrishnan}, {and} \bibinfo{person}{Terri Hedgpeth}.} \bibinfo{year}{2012}\natexlab{}.
\newblock \showarticletitle{Accessible Skimming: Faster Screen Reading of Web Pages}. In \bibinfo{booktitle}{\emph{Proceedings of the 25th Annual ACM Symposium on User Interface Software and Technology}} (Cambridge, Massachusetts, USA) \emph{(\bibinfo{series}{UIST '12})}. \bibinfo{publisher}{Association for Computing Machinery}, \bibinfo{address}{New York, NY, USA}, \bibinfo{pages}{367–378}.
\newblock
\showISBNx{9781450315807}
\urldef\tempurl%
\url{https://doi.org/10.1145/2380116.2380164}
\showDOI{\tempurl}


\bibitem[Blanco et~al\mbox{.}(2022)]%
        {blanco_olli_2022}
\bibfield{author}{\bibinfo{person}{Matthew Blanco}, \bibinfo{person}{Jonathan Zong}, {and} \bibinfo{person}{Arvind Satyanarayan}.} \bibinfo{year}{2022}\natexlab{}.
\newblock \showarticletitle{Olli: {An} {Extensible} {Visualization} {Library} for {Screen} {Reader} {Accessibility}}. In \bibinfo{booktitle}{\emph{{IEEE} {VIS} {Posters}}}.
\newblock
\urldef\tempurl%
\url{http://vis.csail.mit.edu/pubs/olli/}
\showURL{%
\tempurl}


\bibitem[Borowski et~al\mbox{.}(2022)]%
        {borowski_varv_2022}
\bibfield{author}{\bibinfo{person}{Marcel Borowski}, \bibinfo{person}{Luke Murray}, \bibinfo{person}{Rolf Bagge}, \bibinfo{person}{Janus~Bager Kristensen}, \bibinfo{person}{Arvind Satyanarayan}, {and} \bibinfo{person}{Clemens~Nylandsted Klokmose}.} \bibinfo{year}{2022}\natexlab{}.
\newblock \showarticletitle{Varv: {Reprogrammable} {Interactive} {Software} as a {Declarative} {Data} {Structure}}. In \bibinfo{booktitle}{\emph{{CHI} {Conference} on {Human} {Factors} in {Computing} {Systems}}}. \bibinfo{publisher}{ACM}, \bibinfo{address}{New Orleans LA USA}, \bibinfo{pages}{1--20}.
\newblock
\showISBNx{978-1-4503-9157-3}
\urldef\tempurl%
\url{https://doi.org/10.1145/3491102.3502064}
\showDOI{\tempurl}


\bibitem[Brehmer and Munzner(2013)]%
        {brehmer_multi-level_2013}
\bibfield{author}{\bibinfo{person}{Matthew Brehmer} {and} \bibinfo{person}{Tamara Munzner}.} \bibinfo{year}{2013}\natexlab{}.
\newblock \showarticletitle{A {Multi}-{Level} {Typology} of {Abstract} {Visualization} {Tasks}}.
\newblock \bibinfo{journal}{\emph{IEEE Transactions on Visualization and Computer Graphics}} \bibinfo{volume}{19}, \bibinfo{number}{12} (\bibinfo{date}{Dec.} \bibinfo{year}{2013}), \bibinfo{pages}{2376--2385}.
\newblock
\showISSN{1077-2626}
\urldef\tempurl%
\url{https://doi.org/10.1109/TVCG.2013.124}
\showDOI{\tempurl}


\bibitem[Cesal(2020)]%
        {cesal_writing_2020}
\bibfield{author}{\bibinfo{person}{Amy Cesal}.} \bibinfo{year}{2020}\natexlab{}.
\newblock \bibinfo{title}{Writing {Alt} {Text} for {Data} {Visualization}}.
\newblock
\newblock
\urldef\tempurl%
\url{https://medium.com/nightingale/writing-alt-text-for-data-visualization-2a218ef43f81}
\showURL{%
\tempurl}


\bibitem[Charmaz(2006)]%
        {charmaz_constructing_2006}
\bibfield{author}{\bibinfo{person}{Kathy Charmaz}.} \bibinfo{year}{2006}\natexlab{}.
\newblock \bibinfo{booktitle}{\emph{Constructing Grounded Theory} (\bibinfo{edition}{1st} ed.)}.
\newblock \bibinfo{publisher}{SAGE Publications Ltd}, \bibinfo{address}{Thousand Oaks, California}.
\newblock
\showISBNx{9780857029140}
\urldef\tempurl%
\url{https://us.sagepub.com/en-us/nam/constructing-grounded-theory/book235960}
\showURL{%
\tempurl}


\bibitem[Chen et~al\mbox{.}(2023)]%
        {chen_why_2023}
\bibfield{author}{\bibinfo{person}{Yiqun~T. Chen}, \bibinfo{person}{Angela D.~R. Smith}, \bibinfo{person}{Katharina Reinecke}, {and} \bibinfo{person}{Alexandra To}.} \bibinfo{year}{2023}\natexlab{}.
\newblock \showarticletitle{Why, when, and from whom: considerations for collecting and reporting race and ethnicity data in {HCI}}. In \bibinfo{booktitle}{\emph{Proceedings of the 2023 {CHI} {Conference} on {Human} {Factors} in {Computing} {Systems}}} \emph{(\bibinfo{series}{{CHI} '23})}. \bibinfo{publisher}{Association for Computing Machinery}, \bibinfo{address}{New York, NY, USA}, \bibinfo{pages}{1--15}.
\newblock
\showISBNx{978-1-4503-9421-5}
\urldef\tempurl%
\url{https://doi.org/10.1145/3544548.3581122}
\showDOI{\tempurl}


\bibitem[Elavsky(2023)]%
        {elavsky_option-driven_2023}
\bibfield{author}{\bibinfo{person}{Frank Elavsky}.} \bibinfo{year}{2023}\natexlab{}.
\newblock \bibinfo{title}{Option-{Driven} {Design}: {Context}, {Tradeoffs}, and {Considerations} for {Accessibility}}.
\newblock
\newblock
\urldef\tempurl%
\url{http://arxiv.org/abs/2304.08748}
\showURL{%
\tempurl}
\newblock
\shownote{arXiv:2304.08748 [cs]}.


\bibitem[Elavsky et~al\mbox{.}(2022)]%
        {elavsky_how_2022}
\bibfield{author}{\bibinfo{person}{Frank Elavsky}, \bibinfo{person}{Cynthia Bennett}, {and} \bibinfo{person}{Dominik Moritz}.} \bibinfo{year}{2022}\natexlab{}.
\newblock \showarticletitle{How accessible is my visualization? {Evaluating} visualization accessibility with {Chartability}}.
\newblock \bibinfo{journal}{\emph{Computer Graphics Forum}} \bibinfo{volume}{41}, \bibinfo{number}{3} (\bibinfo{date}{June} \bibinfo{year}{2022}), \bibinfo{pages}{57--70}.
\newblock
\showISSN{0167-7055, 1467-8659}
\urldef\tempurl%
\url{https://doi.org/10.1111/cgf.14522}
\showDOI{\tempurl}


\bibitem[Frøkjær and Hornbæk(2005)]%
        {frokjaer_cooperative_2005}
\bibfield{author}{\bibinfo{person}{Erik Frøkjær} {and} \bibinfo{person}{Kasper Hornbæk}.} \bibinfo{year}{2005}\natexlab{}.
\newblock \showarticletitle{Cooperative usability testing: complementing usability tests with user-supported interpretation sessions}. In \bibinfo{booktitle}{\emph{{CHI} '05 {Extended} {Abstracts} on {Human} {Factors} in {Computing} {Systems}}} \emph{(\bibinfo{series}{{CHI} {EA} '05})}. \bibinfo{publisher}{Association for Computing Machinery}, \bibinfo{address}{New York, NY, USA}, \bibinfo{pages}{1383--1386}.
\newblock
\showISBNx{978-1-59593-002-6}
\urldef\tempurl%
\url{https://doi.org/10.1145/1056808.1056922}
\showDOI{\tempurl}


\bibitem[Hamraie(2013)]%
        {hamraie_designing_2013}
\bibfield{author}{\bibinfo{person}{Aimi Hamraie}.} \bibinfo{year}{2013}\natexlab{}.
\newblock \showarticletitle{Designing {Collective} {Access}: {A} {Feminist} {Disability} {Theory} of {Universal} {Design}}.
\newblock \bibinfo{journal}{\emph{Disability Studies Quarterly}} \bibinfo{volume}{33}, \bibinfo{number}{4} (\bibinfo{date}{Sept.} \bibinfo{year}{2013}).
\newblock
\showISSN{2159-8371}
\urldef\tempurl%
\url{https://doi.org/10.18061/dsq.v33i4.3871}
\showDOI{\tempurl}
\newblock
\shownote{Number: 4}.


\bibitem[Joyner et~al\mbox{.}(2022)]%
        {joyner_visualization_2022}
\bibfield{author}{\bibinfo{person}{Shakila Cherise~S Joyner}, \bibinfo{person}{Amalia Riegelhuth}, \bibinfo{person}{Kathleen Garrity}, \bibinfo{person}{Yea-Seul Kim}, {and} \bibinfo{person}{Nam~Wook Kim}.} \bibinfo{year}{2022}\natexlab{}.
\newblock \showarticletitle{Visualization {Accessibility} in the {Wild}: {Challenges} {Faced} by {Visualization} {Designers}}. In \bibinfo{booktitle}{\emph{{CHI} {Conference} on {Human} {Factors} in {Computing} {Systems}}}. \bibinfo{publisher}{ACM}, \bibinfo{address}{New Orleans LA USA}, \bibinfo{pages}{1--19}.
\newblock
\showISBNx{978-1-4503-9157-3}
\urldef\tempurl%
\url{https://doi.org/10.1145/3491102.3517630}
\showDOI{\tempurl}


\bibitem[Jung et~al\mbox{.}(2022)]%
        {jung_communicating_2021}
\bibfield{author}{\bibinfo{person}{Crescentia Jung}, \bibinfo{person}{Shubham Mehta}, \bibinfo{person}{Atharva Kulkarni}, \bibinfo{person}{Yuhang Zhao}, {and} \bibinfo{person}{Yea-Seul Kim}.} \bibinfo{year}{2022}\natexlab{}.
\newblock \bibinfo{title}{Communicating {Visualizations} without {Visuals}: {Investigation} of {Visualization} {Alternative} {Text} for {People} with {Visual} {Impairments}}.
\newblock
\newblock
\urldef\tempurl%
\url{https://doi.org/10.1109/TVCG.2021.3114846}
\showDOI{\tempurl}


\bibitem[Kay and Goldberg(1977)]%
        {kay_personal_1977}
\bibfield{author}{\bibinfo{person}{A. Kay} {and} \bibinfo{person}{A. Goldberg}.} \bibinfo{year}{1977}\natexlab{}.
\newblock \showarticletitle{Personal {Dynamic} {Media}}.
\newblock \bibinfo{journal}{\emph{Computer}} \bibinfo{volume}{10}, \bibinfo{number}{3} (\bibinfo{date}{March} \bibinfo{year}{1977}), \bibinfo{pages}{31--41}.
\newblock
\showISSN{1558-0814}
\urldef\tempurl%
\url{https://doi.org/10.1109/C-M.1977.217672}
\showDOI{\tempurl}
\newblock
\shownote{Conference Name: Computer}.


\bibitem[Kim et~al\mbox{.}(2021)]%
        {kim_accessible_2021}
\bibfield{author}{\bibinfo{person}{N.~W. Kim}, \bibinfo{person}{S.~C. Joyner}, \bibinfo{person}{A. Riegelhuth}, {and} \bibinfo{person}{Y. Kim}.} \bibinfo{year}{2021}\natexlab{}.
\newblock \showarticletitle{Accessible {Visualization}: {Design} {Space}, {Opportunities}, and {Challenges}}.
\newblock \bibinfo{journal}{\emph{Computer Graphics Forum}} \bibinfo{volume}{40}, \bibinfo{number}{3} (\bibinfo{date}{June} \bibinfo{year}{2021}), \bibinfo{pages}{173--188}.
\newblock
\showISSN{0167-7055, 1467-8659}
\urldef\tempurl%
\url{https://doi.org/10.1111/cgf.14298}
\showDOI{\tempurl}


\bibitem[Klokmose et~al\mbox{.}(2015)]%
        {klokmose_webstrates_2015}
\bibfield{author}{\bibinfo{person}{Clemens~N. Klokmose}, \bibinfo{person}{James~R. Eagan}, \bibinfo{person}{Siemen Baader}, \bibinfo{person}{Wendy Mackay}, {and} \bibinfo{person}{Michel Beaudouin-Lafon}.} \bibinfo{year}{2015}\natexlab{}.
\newblock \showarticletitle{Webstrates: {Shareable} {Dynamic} {Media}}. In \bibinfo{booktitle}{\emph{Proceedings of the 28th {Annual} {ACM} {Symposium} on {User} {Interface} {Software} \& {Technology}}}. \bibinfo{publisher}{ACM}, \bibinfo{address}{Charlotte NC USA}, \bibinfo{pages}{280--290}.
\newblock
\showISBNx{978-1-4503-3779-3}
\urldef\tempurl%
\url{https://doi.org/10.1145/2807442.2807446}
\showDOI{\tempurl}


\bibitem[Lundgard and Satyanarayan(2022)]%
        {lundgard_accessible_2022}
\bibfield{author}{\bibinfo{person}{Alan Lundgard} {and} \bibinfo{person}{Arvind Satyanarayan}.} \bibinfo{year}{2022}\natexlab{}.
\newblock \showarticletitle{Accessible {Visualization} via {Natural} {Language} {Descriptions}: {A} {Four}-{Level} {Model} of {Semantic} {Content}}.
\newblock \bibinfo{journal}{\emph{IEEE Transactions on Visualization and Computer Graphics}} \bibinfo{volume}{28}, \bibinfo{number}{1} (\bibinfo{date}{Jan.} \bibinfo{year}{2022}), \bibinfo{pages}{1073--1083}.
\newblock
\showISSN{1077-2626, 1941-0506, 2160-9306}
\urldef\tempurl%
\url{https://doi.org/10.1109/TVCG.2021.3114770}
\showDOI{\tempurl}


\bibitem[Morris et~al\mbox{.}(2018)]%
        {morris_rich_2018}
\bibfield{author}{\bibinfo{person}{Meredith~Ringel Morris}, \bibinfo{person}{Jazette Johnson}, \bibinfo{person}{Cynthia~L. Bennett}, {and} \bibinfo{person}{Edward Cutrell}.} \bibinfo{year}{2018}\natexlab{}.
\newblock \showarticletitle{Rich {Representations} of {Visual} {Content} for {Screen} {Reader} {Users}}. In \bibinfo{booktitle}{\emph{Proceedings of the 2018 {CHI} {Conference} on {Human} {Factors} in {Computing} {Systems}}}. \bibinfo{publisher}{ACM}, \bibinfo{address}{Montreal QC Canada}, \bibinfo{pages}{1--11}.
\newblock
\showISBNx{978-1-4503-5620-6}
\urldef\tempurl%
\url{https://doi.org/10.1145/3173574.3173633}
\showDOI{\tempurl}


\bibitem[Peck et~al\mbox{.}(2019)]%
        {peck_data_2019}
\bibfield{author}{\bibinfo{person}{Evan~M. Peck}, \bibinfo{person}{Sofia~E. Ayuso}, {and} \bibinfo{person}{Omar El-Etr}.} \bibinfo{year}{2019}\natexlab{}.
\newblock \showarticletitle{Data is {Personal}: {Attitudes} and {Perceptions} of {Data} {Visualization} in {Rural} {Pennsylvania}}. In \bibinfo{booktitle}{\emph{Proceedings of the 2019 {CHI} {Conference} on {Human} {Factors} in {Computing} {Systems}}} \emph{(\bibinfo{series}{{CHI} '19})}. \bibinfo{publisher}{Association for Computing Machinery}, \bibinfo{address}{New York, NY, USA}, \bibinfo{pages}{1--12}.
\newblock
\showISBNx{978-1-4503-5970-2}
\urldef\tempurl%
\url{https://doi.org/10.1145/3290605.3300474}
\showDOI{\tempurl}


\bibitem[Pirolli and Card(1999)]%
        {pirolli_information_1999}
\bibfield{author}{\bibinfo{person}{Peter Pirolli} {and} \bibinfo{person}{Stuart Card}.} \bibinfo{year}{1999}\natexlab{}.
\newblock \showarticletitle{Information foraging}.
\newblock \bibinfo{journal}{\emph{Psychological Review}} \bibinfo{volume}{106}, \bibinfo{number}{4} (\bibinfo{year}{1999}), \bibinfo{pages}{643--675}.
\newblock
\showISSN{1939-1471(Electronic),0033-295X(Print)}
\urldef\tempurl%
\url{https://doi.org/10.1037/0033-295X.106.4.643}
\showDOI{\tempurl}
\newblock
\shownote{Place: US Publisher: American Psychological Association}.


\bibitem[Potluri et~al\mbox{.}(2021)]%
        {potluri_examining_2021}
\bibfield{author}{\bibinfo{person}{Venkatesh Potluri}, \bibinfo{person}{Tadashi~E Grindeland}, \bibinfo{person}{Jon~E. Froehlich}, {and} \bibinfo{person}{Jennifer Mankoff}.} \bibinfo{year}{2021}\natexlab{}.
\newblock \showarticletitle{Examining {Visual} {Semantic} {Understanding} in {Blind} and {Low}-{Vision} {Technology} {Users}}. In \bibinfo{booktitle}{\emph{Proceedings of the 2021 {CHI} {Conference} on {Human} {Factors} in {Computing} {Systems}}}. \bibinfo{publisher}{ACM}, \bibinfo{address}{Yokohama Japan}, \bibinfo{pages}{1--14}.
\newblock
\showISBNx{978-1-4503-8096-6}
\urldef\tempurl%
\url{https://doi.org/10.1145/3411764.3445040}
\showDOI{\tempurl}


\bibitem[Saunders et~al\mbox{.}(2015)]%
        {saunders_anonymising_2015}
\bibfield{author}{\bibinfo{person}{Benjamin Saunders}, \bibinfo{person}{Jenny Kitzinger}, {and} \bibinfo{person}{Celia Kitzinger}.} \bibinfo{year}{2015}\natexlab{}.
\newblock \showarticletitle{Anonymising interview data: challenges and compromise in practice}.
\newblock \bibinfo{journal}{\emph{Qualitative Research}} \bibinfo{volume}{15}, \bibinfo{number}{5} (\bibinfo{date}{Oct.} \bibinfo{year}{2015}), \bibinfo{pages}{616--632}.
\newblock
\showISSN{1468-7941}
\urldef\tempurl%
\url{https://doi.org/10.1177/1468794114550439}
\showDOI{\tempurl}
\newblock
\shownote{Publisher: SAGE Publications}.


\bibitem[Sharif et~al\mbox{.}(2021)]%
        {sharif_understanding_2021}
\bibfield{author}{\bibinfo{person}{Ather Sharif}, \bibinfo{person}{Sanjana~Shivani Chintalapati}, \bibinfo{person}{Jacob~O. Wobbrock}, {and} \bibinfo{person}{Katharina Reinecke}.} \bibinfo{year}{2021}\natexlab{}.
\newblock \showarticletitle{Understanding {Screen}-{Reader} {Users}’ {Experiences} with {Online} {Data} {Visualizations}}. In \bibinfo{booktitle}{\emph{Proceedings of the 23rd {International} {ACM} {SIGACCESS} {Conference} on {Computers} and {Accessibility}}} \emph{(\bibinfo{series}{{ASSETS} '21})}. \bibinfo{publisher}{Association for Computing Machinery}, \bibinfo{address}{New York, NY, USA}, \bibinfo{pages}{1--16}.
\newblock
\showISBNx{978-1-4503-8306-6}
\urldef\tempurl%
\url{https://doi.org/10.1145/3441852.3471202}
\showDOI{\tempurl}


\bibitem[Shneiderman(2000)]%
        {shneiderman_eyes_2000}
\bibfield{author}{\bibinfo{person}{Ben Shneiderman}.} \bibinfo{year}{2000}\natexlab{}.
\newblock \showarticletitle{The {Eyes} {Have} {It}: {A} {Task} by {Data} {Type} {Taxonomy} for {Information} {Visualizations}}.
\newblock \bibinfo{journal}{\emph{IEEE Symposium on Visual Languages, Proceedings}} (\bibinfo{date}{March} \bibinfo{year}{2000}).
\newblock
\urldef\tempurl%
\url{https://doi.org/10.1109/VL.1996.545307}
\showDOI{\tempurl}


\bibitem[Smith and Preiser(2011)]%
        {smith_universal_2011}
\bibfield{author}{\bibinfo{person}{Korydon~H. Smith} {and} \bibinfo{person}{Wolfgang F.~E. Preiser}.} \bibinfo{year}{2011}\natexlab{}.
\newblock \bibinfo{booktitle}{\emph{Universal design handbook} (\bibinfo{edition}{2nd} ed.)}.
\newblock \bibinfo{publisher}{McGraw-Hill}, \bibinfo{address}{New York}.
\newblock
\showISBNx{978-0-07-162922-5}


\bibitem[Stangl et~al\mbox{.}(2020)]%
        {stangl_person_2020}
\bibfield{author}{\bibinfo{person}{Abigale Stangl}, \bibinfo{person}{Meredith~Ringel Morris}, {and} \bibinfo{person}{Danna Gurari}.} \bibinfo{year}{2020}\natexlab{}.
\newblock \showarticletitle{"{Person}, {Shoes}, {Tree}. {Is} the {Person} {Naked}?" {What} {People} with {Vision} {Impairments} {Want} in {Image} {Descriptions}}. In \bibinfo{booktitle}{\emph{Proceedings of the 2020 {CHI} {Conference} on {Human} {Factors} in {Computing} {Systems}}} \emph{(\bibinfo{series}{{CHI} '20})}. \bibinfo{publisher}{Association for Computing Machinery}, \bibinfo{address}{New York, NY, USA}, \bibinfo{pages}{1--13}.
\newblock
\showISBNx{978-1-4503-6708-0}
\urldef\tempurl%
\url{https://doi.org/10.1145/3313831.3376404}
\showDOI{\tempurl}


\bibitem[Stofer and Che(2014)]%
        {stofer_comparing_2014}
\bibfield{author}{\bibinfo{person}{Kathryn Stofer} {and} \bibinfo{person}{Xuan Che}.} \bibinfo{year}{2014}\natexlab{}.
\newblock \showarticletitle{Comparing {Experts} and {Novices} on {Scaffolded} {Data} {Visualizations} using {Eye}-tracking}.
\newblock \bibinfo{journal}{\emph{Journal of Eye Movement Research}} \bibinfo{volume}{7}, \bibinfo{number}{5} (\bibinfo{date}{Dec.} \bibinfo{year}{2014}).
\newblock
\showISSN{1995-8692}
\urldef\tempurl%
\url{https://doi.org/10.16910/jemr.7.5.2}
\showDOI{\tempurl}
\newblock
\shownote{Number: 5}.


\bibitem[Williamson(2012)]%
        {williamson_getting_2012}
\bibfield{author}{\bibinfo{person}{Bess Williamson}.} \bibinfo{year}{2012}\natexlab{}.
\newblock \showarticletitle{Getting a {Grip}: {Disability} in {American} {Industrial} {Design} of the {Late} {Twentieth} {Century}}.
\newblock \bibinfo{journal}{\emph{Winterthur Portfolio}} \bibinfo{volume}{46}, \bibinfo{number}{4} (\bibinfo{date}{Dec.} \bibinfo{year}{2012}), \bibinfo{pages}{213--236}.
\newblock
\showISSN{0084-0416}
\urldef\tempurl%
\url{https://doi.org/10.1086/669668}
\showDOI{\tempurl}
\newblock
\shownote{Publisher: The University of Chicago Press}.


\bibitem[Wobbrock et~al\mbox{.}(2011)]%
        {wobbrock_ability-based_2011}
\bibfield{author}{\bibinfo{person}{Jacob~O. Wobbrock}, \bibinfo{person}{Shaun~K. Kane}, \bibinfo{person}{Krzysztof~Z. Gajos}, \bibinfo{person}{Susumu Harada}, {and} \bibinfo{person}{Jon Froehlich}.} \bibinfo{year}{2011}\natexlab{}.
\newblock \showarticletitle{Ability-{Based} {Design}: {Concept}, {Principles} and {Examples}}.
\newblock \bibinfo{journal}{\emph{ACM Transactions on Accessible Computing}} \bibinfo{volume}{3}, \bibinfo{number}{3} (\bibinfo{date}{April} \bibinfo{year}{2011}), \bibinfo{pages}{1--27}.
\newblock
\showISSN{1936-7228, 1936-7236}
\urldef\tempurl%
\url{https://doi.org/10.1145/1952383.1952384}
\showDOI{\tempurl}


\bibitem[Zong et~al\mbox{.}(2022)]%
        {zong_rich_2022}
\bibfield{author}{\bibinfo{person}{Jonathan Zong}, \bibinfo{person}{Crystal Lee}, \bibinfo{person}{Alan Lundgard}, \bibinfo{person}{JiWoong Jang}, \bibinfo{person}{Daniel Hajas}, {and} \bibinfo{person}{Arvind Satyanarayan}.} \bibinfo{year}{2022}\natexlab{}.
\newblock \showarticletitle{Rich {Screen} {Reader} {Experiences} for {Accessible} {Data} {Visualization}}.
\newblock  (\bibinfo{date}{June} \bibinfo{year}{2022}).
\newblock
\urldef\tempurl%
\url{https://doi.org/10.1111/cgf.14519}
\showDOI{\tempurl}


\end{thebibliography}

\end{document}